\documentclass[reprint,amsmath,amssymb,aps,prl]{revtex4-2}

\usepackage{graphicx}
\usepackage{dcolumn}
\usepackage{bm}

\begin{document}

\title{Eigenpulses of dispersive time-varying media}

\author{S. A. R. Horsley}
\email{s.horsley@exeter.ac.uk}
\affiliation{School of Physics and Astronomy, University of Exeter, Stocker Road, Exeter, UK, EX4 4QL}
\author{E. Galiffi}%
\email{egaliffi@gc.cuny.edu}
\affiliation{Photonics Initiative, Advanced Science Research Center, City University of New York, 10027 New York, USA}
\author{Y.-T. Wang}
\affiliation{Photonics Initiative, Advanced Science Research Center, City University of New York, 10027 New York, USA}
%
%
\begin{abstract}
We develop a compact theory that can be applied to a variety of time-varying dispersive materials. The continuous wave reflection and transmission coefficients are replaced with equivalent operator expressions. In addition to comparing this approach to existing numerical and analytical techniques, we find that the eigenfunctions of these operators represent pulses that do not change their spectra after interaction with the time-varying, dispersive material. In addition, the poles of these operators represent the non--time harmonic bound states of the system.
\end{abstract}

\maketitle
%
%
Typical electromagnetic parameters (e.g. refractive index, or impedance) are constant in time, deriving from both the dispersive response and \emph{spatial} arrangement of the constituent atoms, or meta--atoms. Designing such composite materials (metamaterials) with specified behaviour has been the subject of intense research for decades, with recent developments including one--way propagation~\cite{segev2021}, PT--symmetric lasing~\cite{ganainy2018}, and a variety of methods for invisibility cloaking~\cite{lee2021}. Yet many fundamental limitations can be overcome if the material properties are additionally structured in \emph{time} as well as space~\cite{galiffi2022b}. For example, a lossless time-varying grating can amplify waves~\cite{galiffi2022a}, something impossible with a static grating.  The Kramers--Kronig relations~\cite{volume5} also restrict the available palette of static materials (limiting e.g. the thickness to bandwidth ratio of an absorber~\cite{rozanov2000}), but do not apply to time--varying media (TVM)~\cite{solis2021}. In addition, TVM exhibit fundamentally new phenomena such as `time-refraction'~\cite{zhou2020,bohn2021b}, `time-reflection'~\cite{morgenthaler1958velocity,mendonca2002,bacot2016time,moussa2022observation}, and `temporal aiming'~\cite{pena2020}, where wave energy can be re--directed in the absence of spatial inhomogeneity.

While many of the properties of TVM are yet to be experimentally explored, one promising platform is the conducting compound Indium Tin Oxide (ITO)~\cite{jaffray2022}. Close to the epsilon--near--zero (ENZ) frequency of ITO, it exhibits a large non-linear susceptibility~\cite{alam2016}, leading to an effective permittivity that is switchable on a sub--picosecond timescale~\cite{bohn2021a}. This rapid switchability has been used to demonstrate time-refraction~\cite{zhou2020,bohn2021b} and time-diffraction~\cite{tirole2022double} as well as to implement both time-varying metasurfaces~\cite{liu2021}, and spectrum-modifying mirrors~\cite{tirole2022}.

As most materials are static, the theoretical tools for treating TVM are not well developed.  For instance, typical TVM are highly dispersive in the frequency regimes where their nonlinearities are strongest, which makes the susceptibility a two--time function. This is difficult to incorporate into existing analytical results or numerical schemes. Several works have found methods to circumvent these difficulties. For instance, Zhou et al.~\cite{zhou2020} introduced an effective non-dispersive time-varying response that was optimized to fit the experimental data. Many current predictions alternatively integrate Maxwell's equations in full over time using e.g. COMSOL Multiphysics~\cite{comsol}, and such an approach can be found in e.g.~\cite{pena2020}. However this becomes computationally intensive for the dispersive structured materials discussed in e.g.~\cite{liu2021,tirole2022}.  

In this Letter we develop a compact semi--analytical theoretical approach that can be applied to a variety of time varying dispersive materials and is particularly well suited to describe the thin film ITO experiments discussed above. We find that the continuous wave reflection and transmission coefficients can be replaced with equivalent operator expressions that are simple to evaluate numerically and act on the spectrum of the incident wave. Although distinct from this work, aspects of our operator-based approach (where e.g. the wave--vector is treated as an operator) appear in the recent extension of Mie theory given by Ptitcyn et al.~\cite{ptitcyn2021}. We find that the eigenfunctions of these operators represent spectra of incident waves that are e.g. unchanged on reflection from a TVM. Furthermore when the eigenvalue of e.g. the reflection operator is zero or infinite, we have respectively the time-varying analogues of a reflectionless medium or a bound state, which we term the `eigenpulses' of the system. In addition, we compare this approach to existing numerical and semi-analytical techniques, evaluating the superior performance of this method in terms of both accuracy and efficiency. 

For a static material, the electric current $\boldsymbol{j}$ is linked to the electric field $\boldsymbol{E}$ through the conductivity $\sigma(t-t')$ that represents the movement of charge in response to the past behaviour of the electric field and depends only on the time \emph{difference} $t-t'$.  When the material is explicitly time-dependent, due to e.g. a pump pulse~\cite{alam2016} (at optical frequencies) or electronic modulation (at radio frequencies), the conductivity can be replaced with a two--time function such that,
\begin{equation}
    \boldsymbol{j}(t)=\int_{-\infty}^{\infty}{\rm d}t'\sigma(t,t-t')\,\boldsymbol{E}(t')\label{eq:D-field}.
\end{equation}
As in the static case, causality requires $\sigma(t,t-t')=0$ when $t'>t$.  We understand $\sigma(t,t-t')$ as the response at time $t$ to the electric field at the earlier time $t'$, where this response is modulated as a function of the observation time $t$. We could also develop the same formalism by taking the first argument of $\sigma$ as $t'$ instead of $t$.  As the two times are related by $t'=t-(t-t')$, our results can be applied to either form of Eq. (\ref{eq:D-field}), with only minor modifications.  We could also equally develop the formalism in terms of the permittivity and/or permeability instead of the conductivity.

Performing a Fourier transform of (\ref{eq:D-field}), $\tilde{\boldsymbol{j}}(\omega)=\int\,{\rm d}t\,\boldsymbol{j}(t)\exp({\rm i}\omega t)$, the frequency dependent current can be written as
\begin{align}
    \tilde{\boldsymbol{j}}(\omega)&=\int_{-\infty}^{\infty}\frac{{\rm d}\omega'}{2\pi}\tilde{\boldsymbol{E}}(\omega')\int_{\infty}^{\infty}{\rm d}t\,\sigma(t,\omega'){\rm e}^{{\rm i}(\omega-\omega')t}\\
    &=\int_{-\infty}^{\infty}{\rm d}\omega'\tilde{\boldsymbol{E}}(\omega')\hat{\sigma}(-{\rm i}\partial_{\omega},\omega')\delta(\omega-\omega')\nonumber\\[5pt]
    &=\hat{\sigma}(-{\rm i}\partial_{\omega},\omega)\tilde{\boldsymbol{E}}(\omega)\label{eq:Dop}
\end{align}
where $\hat{\sigma}$ is the operator obtained by replacing the first argument $t'$ with the operator $-{\rm i}\partial_{\omega}$.  In the final line of (\ref{eq:Dop}), all derivatives $\partial_\omega$ within $\hat{\sigma}$ must be ordered such that they appear to the \emph{left} of all the frequency dependence of $\hat{\sigma}$~\footnote{The opposite (normal) operator ordering (where the $\partial_\omega$ sit on the \emph{right} of the factors of frequency in $\hat{\sigma}$) is required if we specify the conductivity in Eq. (\ref{eq:D-field}) as $\sigma(t',t-t')$.}.  This prescription is reminiscent of the anti--normal operator ordering adopted in quantum mechanics~\cite{shewell1959}.  To use Eq. (\ref{eq:Dop}) we write the operator as e.g. $\hat{\sigma}=\sum_{n}a_{n}(-{\rm i}\partial_{\omega})b_{n}(\omega)$.  The derivative $\partial_{\omega}$ is numerically constructed as an $N\times N$ matrix acting on $N$ frequency points, via the finite difference approximation or a Fourier transform.  The operators $a_{n}$ are then evaluated as matrix valued functions, and $b_n(\omega)$ is a diagonal matrix.  This idea of using an operator valued function is similar to the exponential function of the Hamiltonian operator used as the time evolution operator in quantum mechanics~\cite{sakuri2018}. 

In this work we assume the magnetic permeability is unity, and use the Drude model with a time-varying plasma frequency $\omega_p$ (see Supplementary Information)
\begin{equation}
    \hat{\sigma}(-{\rm i}\partial_{\omega},\omega)=\omega_p^2(-{\rm i}\partial_{\omega})\frac{{\rm i}\epsilon_0}{\omega+{\rm i}\gamma},\label{eq:drude_model}
\end{equation}
where we have imposed the aforementioned anti--normal ordering, and $1/\gamma$ is the collision time.  Note that throughout this work we use the symbol $\omega_0=\omega_p(-\infty)$, i.e. the plasma frequency before the time variation. 

\begin{figure}
    \centering
    \includegraphics[width=\columnwidth]{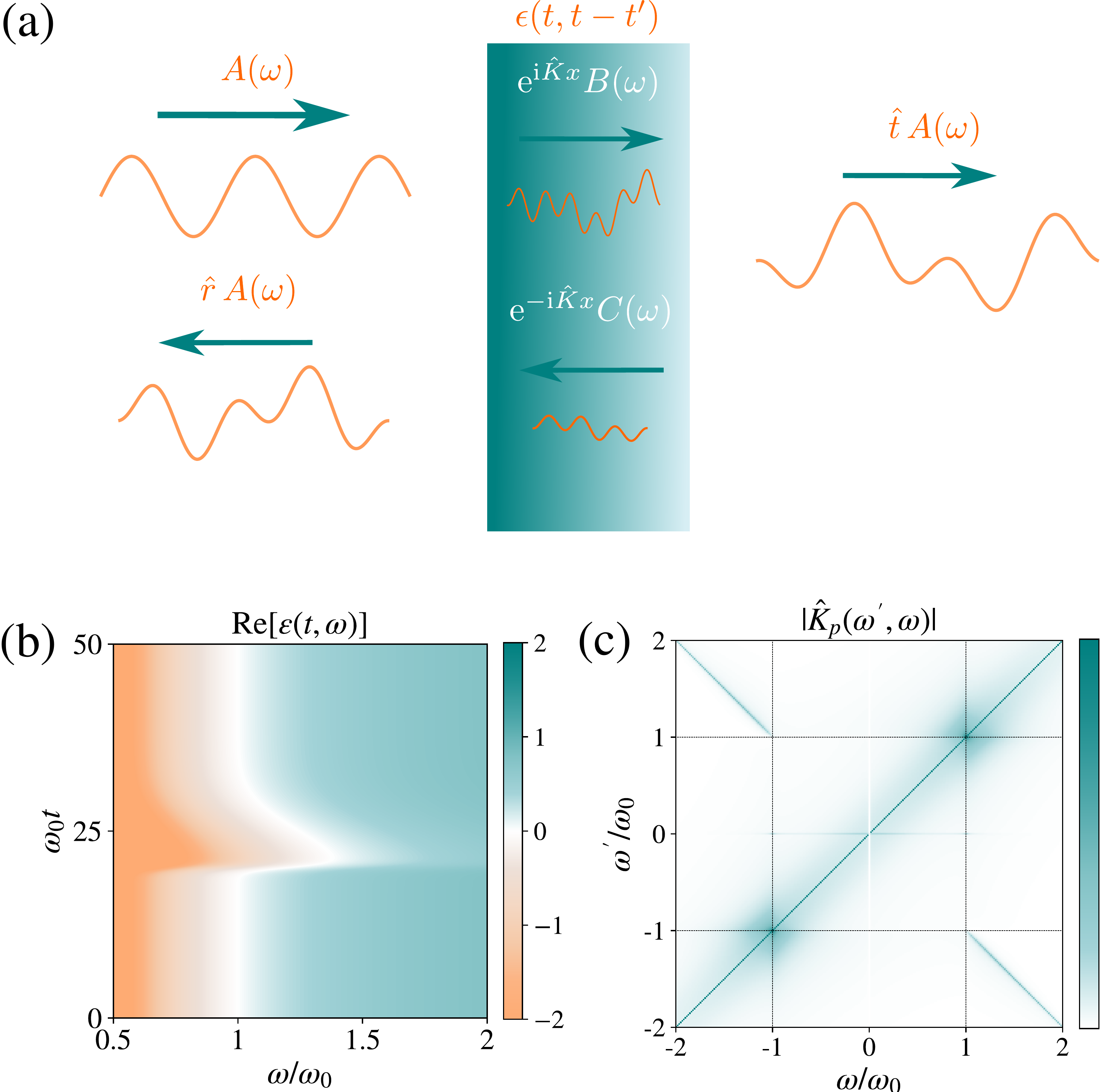}
    \caption{Scattering from a time varying dispersive medium. (a) Our theory treats this as a generalization of a time independent problem, with operators $\hat{r}$ and $\hat{t}$ replacing the usual reflection and transmission coefficients. (b) Here we assume the Drude model permittivity (\ref{eq:drude_model}) with an asymmetric time variation of the plasma frequency, shown here in white.  (c) The operator $\hat{K}_{p}$, determines the spatial evolution of the spectral content of the wave via Eq. (\ref{eq:simple_case}).  For the time dependence shown in panel (b), the operator causes a spectral broadening (the smearing around $\omega=\omega'$, with this largest close to the plasma frequency) and reversal of propagation direction (the line around $\omega=-\omega'$).}
    \label{fig:example1}
\end{figure}

The simplest application of Eq. (\ref{eq:Dop}) is where the medium is homogeneous, propagation is along $x$, and the field is polarized such that $\tilde{\boldsymbol{H}}=H\boldsymbol{e}_{z}$.  In this case Maxwell's equations become $\partial_{x}^2H+\hat{K}_{p}^2 H=0$, the solutions to which are
\begin{equation}
    H(x,\omega)=\exp(\pm{\rm i}\hat{K}_p x)\,H_0(0,\omega)\label{eq:simple_case}
\end{equation}
where $k_0=\omega/c$ and $\hat{K}_p$ is the matrix square root of $\hat{K}^2_p={\rm i}\eta_0\hat{\sigma}k_0+k_0^2$~\footnote{It is important to determine which sign of the matrix square root $\pm\hat{K}_p$ describes right--going waves.  We take right--going waves to be those where the eigenvalues have a positive imaginary part.} and $H(0,\omega)$ is the Fourier amplitude of the wave at $x=0$, which we are free to choose.  The quantity $\eta_0=\sqrt{\mu_0/\epsilon_0}$ is the impedance of free space.  Although the solutions (\ref{eq:simple_case}) have the appearance of plane waves, the operator $\exp(\pm{\rm i}\hat{K}_{p}x)$ modifies the spectral content of the wave as the observation point $x$ is changed, describing the reshaping of the pulse during propagation.  Eq. (\ref{eq:simple_case}) shows that those spectra $H(0,\omega)$ that are eigenfunctions of $\hat{K}_p$ with eigenvalue $\lambda$ have a plane wave spatial dependence $\exp(\pm{\rm i}\lambda x)$ and retain the same spectrum during propagation, despite the time variation of the material parameters.  Note also that Eq. (\ref{eq:simple_case}) is similar to the aforementioned time evolution operator in quantum mechanics, where a state $|\psi\rangle$ evolves in time as $|\psi(t)\rangle=\exp(-{\rm i}\hat{H} t/\hbar)|\psi(0)\rangle$, where $\hat{H}$ is the Hamiltonian operator.  Fig.~\ref{fig:example1} illustrates the basic idea of our formalism, and gives the form of the operator $\hat{K}_{p}$ for a typical time variation of the material parameters (\ref{eq:drude_model}) (See SI for further details).
%
%
%
%
\paragraph{Fresnel coefficients for a dispersive, time--varying interface:}
Consider a pulse incident from vacuum onto a dispersive time varying half--space ($x>0$).  Using the operator formalism described above, we calculate the reflection $\hat{r}$ and transmission $\hat{t}$ operators for an incident pulse.  Just as for static materials, this calculation must be done separately for transverse electric (TE) and transverse magnetic (TM) polarizations.

Assuming incidence in the $x$--$y$ plane with in--plane wave--vector $\boldsymbol{k}_{\parallel}$, TE polarized waves have an electric field $\tilde{\boldsymbol{E}}=E\boldsymbol{e}_{z}$ obeying the operator Helmholtz equation $\partial_{x}^{2} E+\hat{K}^2_{s}E=0$ where $\hat{K}_{s}^{2}={\rm i}\eta_0 k_0\hat{\sigma}+k_0^2-k_{\parallel}^2$.  Inside the TVM ($x>0$) the solution is given by Eq. (\ref{eq:simple_case}), $E(x>0,\omega)=\exp({\rm i}\hat{K}_{s}x)C_s(\omega)$.  Meanwhile, on the entrance side the field is a sum of plane waves for each frequency $E(x<0,\omega)=A_s(\omega)\exp({\rm i}k_x x)+B_s(\omega)\exp(-{\rm i}k_x x)$ where $k_x=[k_0^2-k_{\parallel}^2]^{1/2}$.  The spatial boundary conditions are the same as for static media, with both electric $E$ and magnetic $\eta_0 H_{y}={\rm i}k_0^{-1}\partial_x E$  fields  continuous across $x=0$.  Substituting the forms of the fields in the respective regions leads to the following reflection and transmission operators
\begin{align}
    \hat{r}_s&=\left(1-\hat{Z}_{s}\right)\left(1+\hat{Z}_{s}\right)^{-1}\nonumber\\
    \hat{t}_{s}&=2\left(1+\hat{Z}_s\right)^{-1}\label{eq:fresnelTE}
\end{align}
where $\hat{Z}_{s}=k_x^{-1}\hat{K}_{s}$, $B_s=\hat{r}_{s}A_s$ and $C_s=\hat{t}_s A_s$.  Eqns. (\ref{eq:fresnelTE}) are the TE Fresnel coefficients~\cite{volume8}, with an operator replacing the usual expression for the wave--vector in the material. 

The derivation is slightly different for TM polarization, revealing the importance of operator ordering in these calculations.  Taking the magnetic field as $\boldsymbol{H}=H\boldsymbol{e}_{z}$, it obeys the operator Helmholtz equation $\partial_{x}^2 H+\hat{K}_{p}^2 H=0$.  The square of the wave--vector is as above $\hat{K}_{p}^{2}={\rm i}\eta_0\hat{\sigma}k_0+k_0^2-k_{\parallel}^2$, differing from the TE expression due to the non--commuting nature of $\hat{\sigma}$ and $\omega$.  Applying the continuity of the magnetic $H$, and electric $E=-{\rm i}[k_0+{\rm i}\eta_0\hat{\sigma}]^{-1}\,\,\partial_{x}\eta_0 H$ fields at the $x=0$ interface leads to the reflection and transmission operators
\begin{align}
    \hat{r}_{p}&=\left(1-\hat{Z}_{p}\right)\left(1+\hat{Z}_{p}\right)^{-1}\nonumber\\
    \hat{t}_{p}&=2\left(1+\hat{Z}_{p}\right)^{-1}\label{eq:fresnelTM}
\end{align}
where $\hat{Z}_{p}=k_x^{-1}[1+{\rm i}\eta_0\hat{\sigma}k_0^{-1}]^{-1}\hat{K}_{p}$.  These are again operator analogues of the TM Fresnel coefficients, although in this case the operator ordering would not be obvious without applying the boundary conditions.  Importantly, at normal incidence the two wave--vector operators differ by a similarity transformation: $\hat{K}_{p}=k_0^{-1}\hat{K}_{s}k_0$.  The two impedance operators are then simply related by $\hat{Z}_{p}=\hat{K}_{s}^{-1}k_0=\hat{Z}_{s}^{-1}$, making the reflection operators (\ref{eq:fresnelTE}--\ref{eq:fresnelTM}) differ by a minus sign $\hat{r}_{s}=-\hat{r}_{p}$ as expected for the two polarizations at normal incidence~\footnote{We define the TM reflection coefficients in terms of the \emph{magnetic} field amplitude, and the TE reflection coefficient in terms of the \emph{electric} field amplitude, hence their difference by a sign at normal incidence.}.

\begin{figure}
    \centering
    \includegraphics[width=\columnwidth]{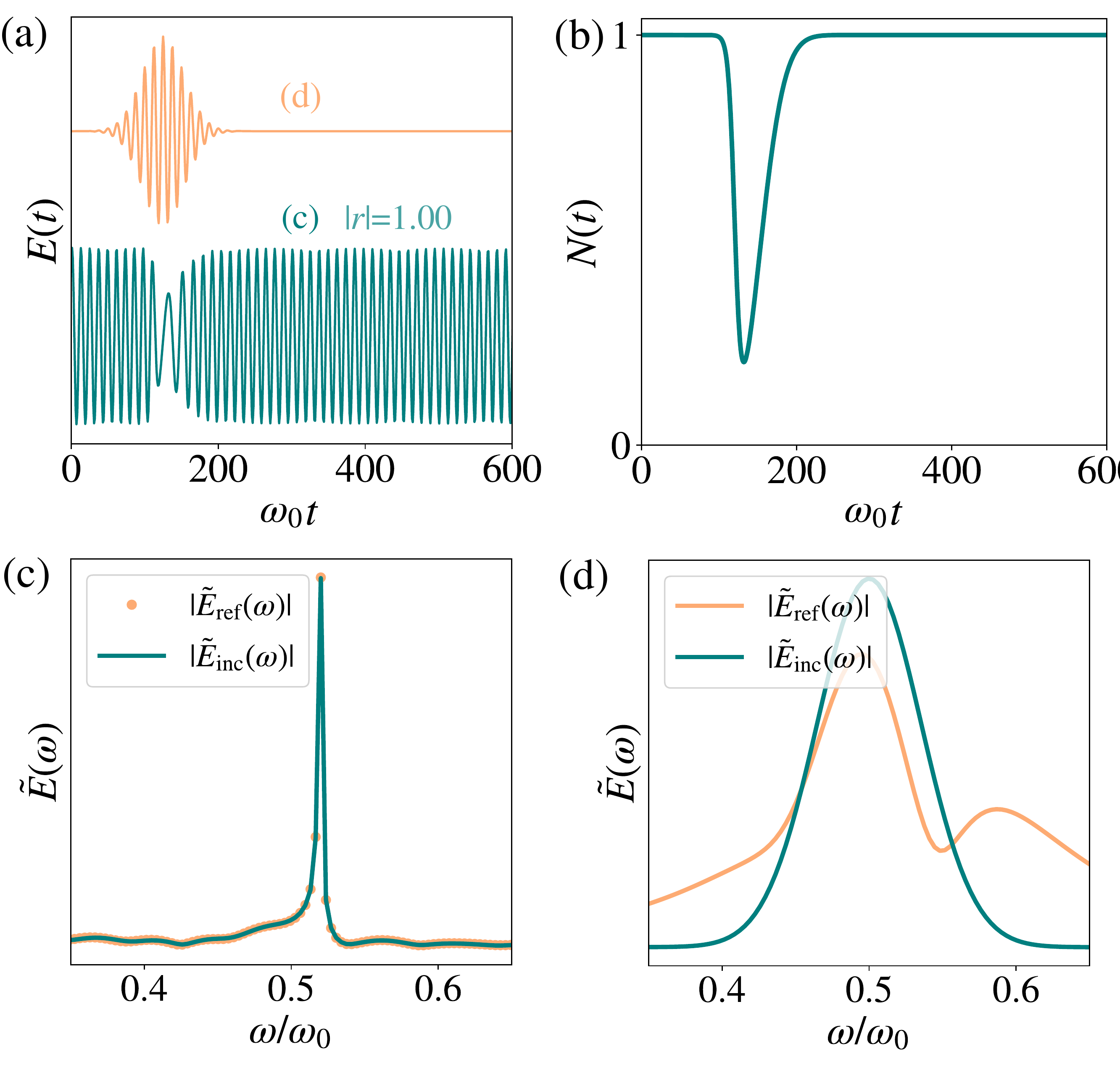}
    \caption{Reflection of an eigenpulse from a time varying dispersive half--space. (a) We compare the reflection of two different incident pulses; a Gaussian pulse (upper curve), and an eigenpulse (lower curve) computed from the reflection operator (\ref{eq:fresnelTM}), with eigenvalue $|r|=1.00$  (the zero level is displaced to aid visualization). (b) Time variation of $N(t)=\omega_p^2(t)/\omega_0^2$ in Eq. (\ref{eq:drude_model}). (c--d) Magnitude of incident and reflected Fourier spectra computed via a numerical integration of Maxwell's equations (See Supplementary Information), for an incident (c) eigenpulse, and (d) Gaussian pulse.}
    \label{fig:example2}
\end{figure}

As discussed above, it is again interesting to examine the eigenvalues and eigenvectors of the reflection and transmission operators.  These reveal that there are pulse spectra (`eigenpulses') that retain an identical spectrum after interaction with the TVM (bar an overall multiplicative factor).  Alternatively, taking a singular value decomposition of the reflection and transmission operators, we can find pulse spectra that are scaled by a set level (the singular value), but have a different spectral content after interaction with the material. We can see from expressions (\ref{eq:fresnelTE}--\ref{eq:fresnelTM}) that, in the case of a single interface, the eigenpulses are the eigenfunctions of the impedance operators $\hat{Z}_{s,p}$ and thus \emph{both} transmitted and reflected spectra are unchanged.  Fig.~\ref{fig:example2} shows a comparison between the reflection of a Gaussian pulse and an eigenpulse from a TVM (plots show the incident field just before the interface).  While the Gaussian pulse is significantly broadened and reshaped by the interaction with the TVM, the eigenpulse reduces in frequency in tandem with the plasma frequency, retaining an identical spectrum upon reflection.  In this case (modulus of eigenvalue $|r|=1$), the eigenpulse is also entirely reflected by the medium, as if it were a mirror. 

In addition, the scattering operators can also exhibit poles.  For example in Eqns. (\ref{eq:fresnelTE}--\ref{eq:fresnelTM}) these poles occur where ${\rm det}(1+\hat{Z}_{s,p})=0$.  The vectors in the null--space of $(1+\hat{Z}_{s,p})$ then represent non--time harmonic modes that are---in this case---confined to the interface of the material.  In the Supplementary Information we find the surface plasmon--like eigenpulses that are confined to the interface of a TVM,
%
%
\paragraph{Time varying layer:}
We can straightforwardly extend this approach to any multilayer and any simple geometry (e.g. a spherical, cylindrical, or ellipsoidal object) that admits an analytic solution to Maxwell's equations in the static case. Broadly speaking, the results for the scattering operators will have an identical form but with an operator replacing the material parameters. To illustrate this in a non--trivial case we calculate the reflection and transmission operators for a slab of thickness $d$, which is relevant to the experiments reported in~\cite{zhou2020,bohn2021a,bohn2021b,tirole2022,tirole2022double}.
%
%
\begin{figure}
    \centering
    \includegraphics[width=\columnwidth]{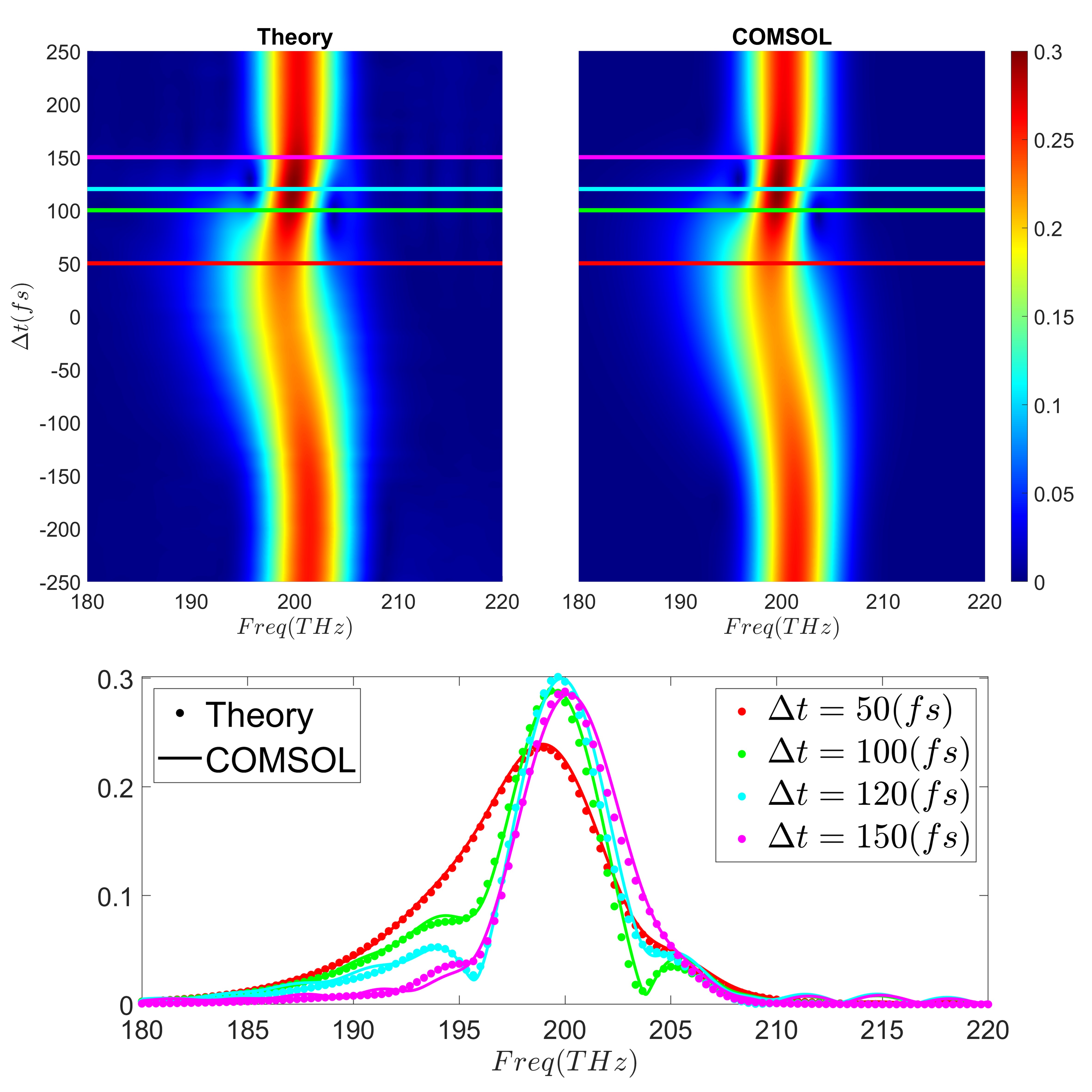}
    \caption{Comparison between the proposed theory and COMSOL multiphysics~\cite{comsol}. The figures demonstrate normalized Fourier transmitted spectra for a time-modulated Drude slab excited by a 45 degree incident probe wave. (See Supplementary Information for more details)}
    \label{fig:example3}
\end{figure}

Assuming TM polarization, the magnetic field within the layer is taken to be of the form
\begin{equation}
    H(0<x<d,\omega)={\rm e}^{{\rm i}\hat{K}_{p}x}C_{p}(\omega)+{\rm e}^{-{\rm i}\hat{K}_{p}x}D_{p}(\omega)
\end{equation}
with the field in the external regions equal to $H(x<0)=\exp({\rm i}k_x x)A_{p}(\omega)+\exp(-{\rm i}k_x x)B_{p}(\omega)$ and $H(x>d)=\exp({\rm i}k_x (x-d))F_{p}(\omega)$.  Imposing the same boundary conditions described above we obtain the reflection and transmission coefficients for the slab
\begin{multline}
    \hat{r}_{\rm slab}=\left[\hat{A}_{+}{\rm e}^{{\rm i}\hat{K}_{p}d}\hat{A}_{-}-\hat{A}_{-}{\rm e}^{-{\rm i}\hat{K}_{p}d}\hat{A}_{+}\right]\\
    \times\left[\hat{A}_{-}{\rm e}^{{\rm i}\hat{K}_{p}d}\hat{A}_{-}-\hat{A}_{+}{\rm e}^{-{\rm i}\hat{K}_{p}d}\hat{A}_{+}\right]^{-1}\label{eq:slab_r}
\end{multline}
and
\begin{equation}
    \hat{t}_{\rm slab}=4\hat{Z}_{p}\left[\hat{A}_{+}{\rm e}^{-{\rm i}\hat{K}_{p}d}\hat{A}_{+}-\hat{A}_{-}{\rm e}^{{\rm i}\hat{K}_{p}d}\hat{A}_{-}\right]^{-1}\label{eq:slab_t}
\end{equation}
where $\hat{A}_{\pm}=1\pm\hat{Z}_{p}$.  Expressions (\ref{eq:slab_r}--\ref{eq:slab_t}) reduce to the familiar reflection and transmission coefficients of a dielectric slab~\cite{volume8} when the operators are replaced with their scalar counterparts.  When $d=0$ the reflection operator (\ref{eq:slab_r}) is identically zero, and the transmission operator (\ref{eq:slab_t}) becomes the identity, as they should.

In Fig.~\ref{fig:example3} we give a comparison between results obtained using COMSOL Multiphysics (see SI), and calculations made using the reflection and transmission operators (\ref{eq:slab_r}--\ref{eq:slab_t}).  We plot the normalized transmitted spectra as a function of pulse delay time $\Delta t$.  As shown in the lower panel of this figure, there is excellent agreement between the finite element calculation and our operator approach. Additional comparisons to an adiabatic multiple-timescale approach used to model past experiments \cite{bohn2021a,tirole2022} are also available in the SI. Importantly, these tests demonstrate the advantage of this method for the efficient modelling of structures that feature extremely subwavelength layers, circumventing the need for expensive numerical calculations.     
%
%
\paragraph{Summary and Conclusions:}  We have developed a compact theoretical approach for treating the problem of scattering from dispersive TVM.  We have shown that our analytic expressions match full wave numerical simulations well.  Although the theory is formally similar to the case of static materials, the TVM parameters are given in terms of operators that depend on both the frequency and frequency derivatives, which must be carefully ordered.  The advantage of our theory is that it is semi--analytical, allowing us to give explicit operator expressions for scattering coefficients from the TVM, and thus determine conditions for e.g. incoming modes that are bound, not reflected, or completely reflected by the material.  We have numerically constructed these operators and found the `eigenpulses' of a time--modulated Drude half--space, numerically verifying that there are input pulse spectra that e.g. reflect as if the TVM was a static mirror.  This approach may be readily extended to other areas of wave physics such as pressure acoustics~\cite{cho2020} and elasticity and may be of interest to those working on TVM as well as multiple scattering, where our reflectionless eigenpulses are analogous to the concept of open scattering channels in disordered media (see e.g.~\cite{choi2011}).
%
%
\begin{acknowledgments}
\paragraph{Acknowledgements:}  SARH acknowledges the Royal Society and TATA for financial support through grant URF\textbackslash R\textbackslash 211033, and thanks Riccardo Sapienza, Euan Hendry, James Capers, and Dean Patient for useful conversations. EG acknowledges funding from the Simons Foundation through a Junior Fellowship of the Simons Society of Fellows (855344,EG). 
\end{acknowledgments}

\appendix

%
%
\bibliography{refs}

\end{document}